\documentclass[twocolumn]{webofc}
 
\usepackage[varg]{txfonts}   
\usepackage{epsfig,graphicx,float,color}
\usepackage{amssymb}
\usepackage{amsmath}
\usepackage[normalem]{ulem}
\begin{document}

\title{At the edge of shape coexistence in the Z=40 region}
%
%

\author{\firstname{E.} \lastname{Maya-Barbecho}\inst{1}\fnsep\thanks{\email{esperanza.maya@alu.uhu.es}} \and
        \firstname{S.} \lastname{Baid}\inst{2}\fnsep\thanks{\email{sbaid@us.es}} \and
        \firstname{J.M.} \lastname{Arias}\inst{2,3}\fnsep\thanks{\email{ariasc@us.es}} \and
        \firstname{J.E.} \lastname{Garc\'{\i}a-Ramos}\inst{1,3}\fnsep\thanks{\email{enrique.ramos@dfaie.uhu.es}}}
      
      \institute{Departamento de  Ciencias Integradas y Centro de Estudios Avanzados en F\'{\i}sica, Matem\'atica y Computaci\'on, Universidad de Huelva, 21071 Huelva, Spain
        \and Departamento de F\'isica At\'omica, Molecular y Nuclear, Facultad de F\'isica\text{,} Universidad de Sevilla, Apartado 1065, E-41080 Sevilla, Spain
        \and Instituto Carlos I de F\'{\i}sica Te\'orica y Computacional,  Universidad de Granada, Fuentenueva s/n, 18071 Granada, Spain }
      
    \abstract{In this contribution, the shape coexistence phenomenon near the proton sub-shell closure at Z=40 is analyzed. Particular emphasis is placed on extracting the nuclear deformation values by examining experimental B(E2) transition rates and observing how the kinematic moment of inertia evolves in these nuclei. Based on the analysis of these two observables, we arrive at the conclusion that the notable effects of shape coexistence at approximately Z=40 are largely diminished in Mo isotopes and are scarcely detectable in Ru isotopes. 
    } 

\maketitle

\section{Introduction}
\label{intro}
Shape coexistence is a phenomenon observed not only in atomic nuclei but also in other systems, such as certain molecules. It is known to occur in various nuclear mass regions, with a particular prevalence around neutron and proton shell closures \cite{heyde11,Garr22}.

In nuclei near a shell closure, the tendency is for them to be spherical, and particle-hole excitations in such regions are energetically costly. However, the energy of these excitations can be significantly reduced due to the enhanced residual interaction brought about by the presence of more effective valence nucleons  \cite{heyde11}. Consequently, these excitations correspond to more deformed states. This effect becomes particularly significant when protons are near a shell closure, while neutrons are situated in the middle of a shell, or vice versa. In this scenario, shape coexistence arises in nuclei, where regular states (0-particle-0-hole) coexist with intruder states (2-particle-2-hole) within the same energy range despite having distinct deformations. In extreme cases, intruder states can even have lower energies than regular states \cite{Garc19}.

It is essential to note that deformation in nuclear structure is not a directly measurable quantity but must be inferred from certain observable data. In this contribution, we will extract information regarding the deformation of isotopes of Sr, Zr, Mo, and Ru. We will then employ this information to explore how the phenomenon of shape coexistence evolves as we move further away from the Z=40 subshell closure. Of particular interest is the examination of the onset of deformation in different isotopic chains and its potential connection with Quantum Phase Transitions (QPTs) \cite{Cejn09}.

\section{The ``crossing'' of configurations}
\label{sec-crossing}
\begin{figure}
\centering
\sidecaption
\includegraphics[width=0.35\textwidth]{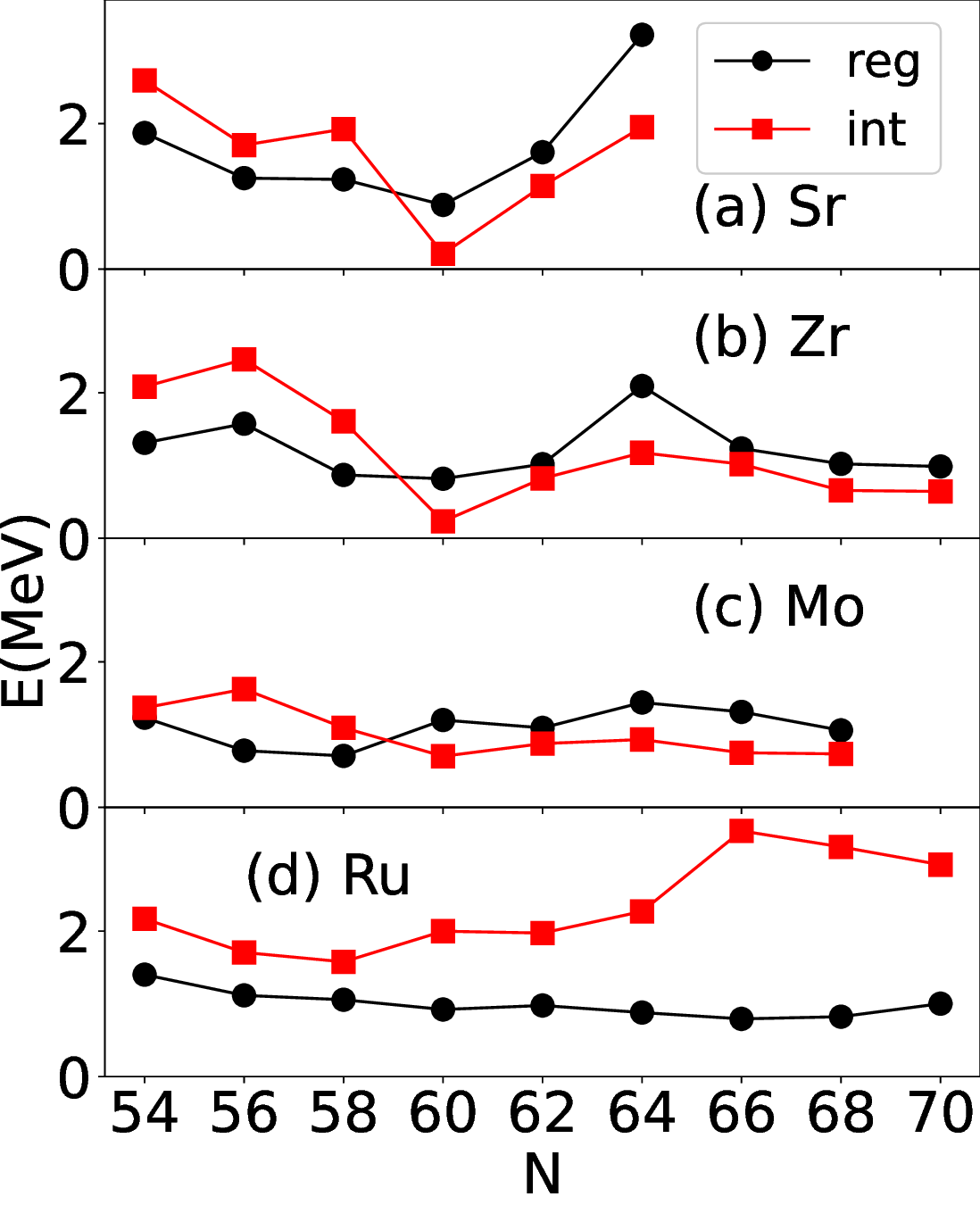}
\caption{Unperturbed regular (black dots and lines) and intruder (red squares and lines) ground state energies for Sr, Zr, Mo, and Ru isotopes as a function of the neutron number.}
\label{fig-crossing}
\end{figure}

Identifying regular and intruder states in a given nucleus is a non-trivial task, and except for the most evident cases, it often requires the use of a theoretical model. One particularly suitable choice is the Interacting Boson Model (IBM) \cite{iach87} with Configuration Mixing (IBM-CM) \cite{duval82}. It incorporates 2p-2h excitations and is valuable for describing the key spectroscopic properties of nuclei providing a microscopic representation of the wave function. With this model, it becomes feasible to determine whether a given state belongs to the regular or the intruder sector.

The IBM-CM has been employed in the study of isotopes of Sr \cite{Maya2022}, Zr \cite{Garc19}, Mo and Ru isotopes \cite{Maya2023}. In Figure \ref{fig-crossing}, we present a visualization of the lowest energy states from both the regular and intruder sectors as extracted from these calculations. Importantly, these energies are computed with the coupling between the two sectors removed, providing a clear separation of pure regular and intruder states.

One striking observation from Figure \ref{fig-crossing} is that regular and intruder states are often quite close in energy, and in certain cases, they even cross paths. These crossings are evident in Sr, Zr, and Mo isotopes, particularly around the neutron number N=60. Notably, even in Ru isotopes, where crossings are absent, the closest approach between regular and intruder states occurs around the same neutron number, N=60. This region is well-known for its rapid evolution of ground state structure and the presence of a QPT at N=60 \cite{Cejn09}. Consequently, it is reasonable to connect the presence of the QPT with the occurrence of crossings between regular and intruder configurations as elucidated in \cite{Gavr19,Gavr22}.

\section{The deformation}
\label{sec-deformation}
\begin{figure}
\centering
\sidecaption
\includegraphics[width=0.4\textwidth]{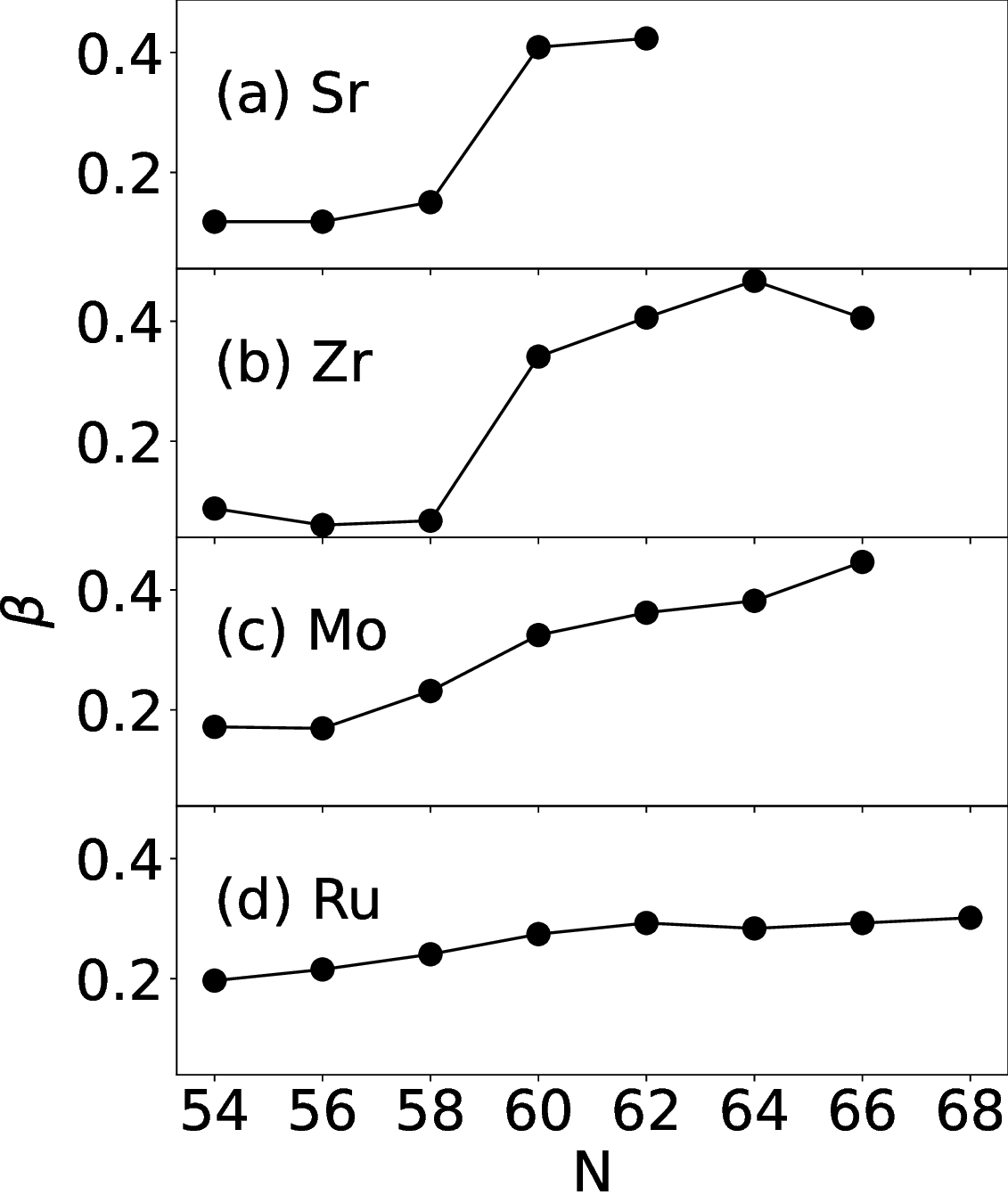}
\caption{Value of the deformation parameter, $\beta$, for the ground state band in Sr, Zr, Mo, and Ru isotopes extracted from $B(E2: 2_1^+\rightarrow 0_1^+)$ as a function of the neutron number.}
\label{fig-deformation}
\end{figure}

To extract information about the quadrupole deformation, $\beta$, for a given band we will rely on the geometrical view of the nucleus extracting the $\beta$ parameter from a known $B(E2)$ value as,
\begin{equation}
\beta=\frac{4\pi\sqrt{B(E2: I_i\rightarrow I_f)}}{3 \langle I_i 0 2 0| I_f 0\rangle Z e R_0^2},
\end{equation} 
where $R_0=1.2 A^{1/3}$ fm and $<.... | ..>$ is the Clebsch-Gordan coefficient and for the case of the yrast band we use $I_i=2$ and $I_f=0$. 

In Figure \ref{fig-deformation}, we present the extracted values of the deformation parameter, $\beta$, for the ground state bands in the four isotopic chains. It is evident that these chains exhibit distinct and contrasting systematics. In the cases of Sr and Zr, there is a clear and rapid onset of deformation at neutron number N=60. For Mo, the increase in deformation is more gradual. In the case of Ru, the value of $\beta$ remains nearly constant. In the instances of Sr and Zr, the sudden increase in deformation arises from the crossing of regular and intruder configurations, as previously shown in Figure \ref{fig-crossing}. Notably, the intruder configuration has a significantly larger deformation, contributing to the rapid change in $\beta$. However, in the case of Mo, although the crossing of regular and intruder configurations exists, the strong interaction between these configurations effectively smooths out the variation in $\beta$ \cite{Maya2023}. This smoothing effect has also been observed in Pt \cite{Garc09}. These observations are further corroborated by the systematics of the two-neutron separation energy \cite{Garc19,Maya2022,Maya2023}. In the language of QPTs, Sr and Zr exhibit a first order Type II QPT, as seen in the behavior of $\beta$ \cite{Gavr19,Gavr22}. Similarly, Mo showcases a second order Type II QPT. In contrast, Ru provides ample evidence of a second-order QPT of Type I \cite{Maya2023}. 

\section{The moment of inertia}
\label{sec-moment}
\begin{figure}
\centering
\sidecaption
\includegraphics[width=0.4\textwidth]{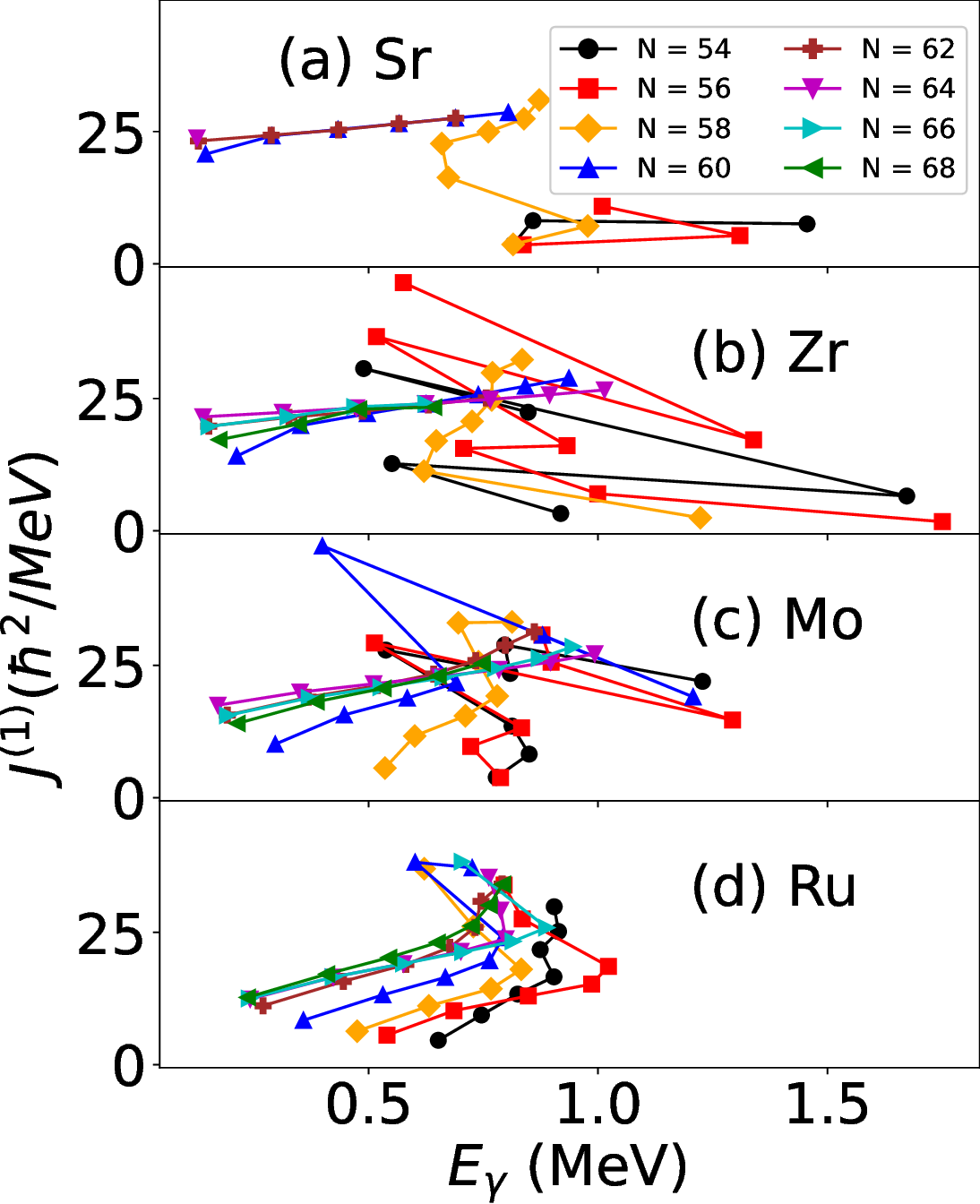}
\caption{Value of the kinematic moment of inertia for Sr, Zr, Mo, and Ru isotopes as a function of the $\gamma$ energy, for different neutron numbers.}
\label{fig-inertia}
\end{figure}

An alternative and valuable approach for gaining insights into the evolving collective structure along the yrast band, with a particular focus on deformation, is to examine the variations in the moment of inertia as one progresses up the band structure. In this context, we will specifically investigate the kinematic moment of inertia, denoted as \textit{J}$^{(1)}$, which is defined as follows \cite{Bohr75},
\begin{equation}
J^{(1)}
=\frac{\hbar^2}{2}\left(\frac{dE}{d\left(J(J+1)\right)}\right)^{-1}
\approx \frac{\hbar^2(2J-1)}{ E_\gamma(J \rightarrow (J-2))} , 
\label{kinetic}
\end{equation}
where $E_\gamma(J \rightarrow (J-2))$ is the energy difference $E(J)-E(J-2)$. 

In constructing the experimental values of \textit{J}$^{(1)}$, we rely on the yrast band energies. However, complexities are encountered at high-spin levels (8$^+$, 10$^+$, and 12$^+$). These complexities arise from the crossing of non-collective states, such as those associated with broken-pair configurations, and the crossing of regular and intruder configurations at specific energy levels. Such crossings can lead to significant variations in the smooth progression of \textit{J}$^{(1)}$ sometimes resulting in a backbending pattern.

In Figure \ref{fig-inertia}, we present the kinematic moment of inertia for the different isotope chains. To aid in visualization, it is helpful to note that a purely rotational system exhibits a constant moment of inertia, while a pure vibrational structure appears as a vertical trend. For Sr isotopes a significant change in behavior is observed from N=60 onward, where the moment of inertia demonstrates a relatively smooth trend, indicative of a rotational structure. However, for N=58, it appears that the lowest angular momenta correspond to a more vibrational-like structure, while the highest angular momenta exhibit rotational characteristics. For N=54-56, it is challenging to draw a definitive conclusion, though the substantial values of E$_\gamma$ (gamma-ray energy) suggest a vibrational system. The results points towards the coexistence of two configurations, with the rotational structure corresponding to the intruder configuration and the vibrational structure to the regular one. For Zr and Mo isotopes, similarly to Sr, the information presented in Figure  \ref{fig-inertia} supports the existence of two different configurations. From N=60 and onwards, there is a clear presence of a rotational structure. However, for N=54-60, the observed backbending suggests the coexistence of two structures within the yrast band. In the Ru isotopes, a relatively consistent behavior is observed across the entire chain. Nevertheless, the highest angular momenta exhibit an upward trend, suggesting the presence of a vibrational structure, while the lowest angular momenta maintain a relatively horizontal trend, typical of a more deformed structure. In summary, the kinematic moment of inertia serves as a suitable observable for studying the coexistence of different configurations. For Sr, Zr, and Mo, the data suggests the presence of two crossing configurations, while in Ru, a single configuration exists, showing more of a rotational nature for lower angular momenta and a more vibrational character for higher angular momenta. 

\section{Conclusions}
\label{sec-conclusions}
This contribution presents a comprehensive study investigating the onset of deformation and its relationship with shape coexistence around the Z=40 region, focusing on the isotope chains $_{38}$Sr, $_{40}$Zr, $_{42}$Mo, and $_{44}$Ru. The study employs experimental data, including the deformation parameter, $\beta$, and kinematic moment of inertia, derived from B(E2) values and excitation energies. The key findings reveal the significance of shape coexistence in explaining energy systematics for Sr and Zr, while Mo exhibits a notable impact from intruder and regular configurations. However, this impact diminishes for Ru isotopes, with N=60 marking a boundary between spherical and deformed nuclei. The study also relates these results to Quantum Phase Transitions (QPTs), with Sr and Zr experiencing first-order Type II QPTs at N=60, Mo showing a second-order Type II QPT, and Ru undergoing a second-order Type I QPT at N=60. This study, complementing \cite{Gavr19, Garc20, Gavr22, Maya2022, Maya2023}, sheds light on the complex interplay between nuclear structure, coexisting configurations, and quantum phase transitions.

\section{Acknowledgements}
This work was partially supported by grants PID2019-104002GB-C21, PID2019-104002GB-C22, PID2022-136228NB-C21, and PID2022-136228NB-C22 funded by MCIN/AEI/10.13039/50110001103 and ``ERDF A way of making Europe''. It was also partially supported by the Consejería de Economía, Conocimiento, Empresas y Universidad de la Junta de Andalucía (Spain) under Group FQM-160, and under Projects No. P20-01247, and No. US-1380840. Resources supporting this work were provided by the CEAFMC and Universidad de Huelva High Performance Computer (HPC@UHU) funded by ERDF/MI\-NE\-CO project UNHU-15CE-2848.

\bibliography{cgs17-mo-ru}

\begin{thebibliography}{13}

\bibitem{heyde11}
K.~Heyde, J.L. Wood, Rev. Mod. Phys. \textbf{83}, 1467 (2011)

\bibitem{Garr22}
P.E. Garrett, M.~Zielińska, E.~Clément, Progress in Particle and Nuclear Physics \textbf{124}, 103931 (2022)

\bibitem{Garc19}
J.E. Garc\'{\i}a-Ramos, K.~Heyde, Phys. Rev. C \textbf{100}, 044315 (2019)

\bibitem{Cejn09}
P.~Cejnar, J.~Jolie, Progress in Particle and Nuclear Physics \textbf{62}, 210  (2009)

\bibitem{iach87}
F.~Iachello, A.~Arima, \emph{{The interacting boson model}} (Cambridge University Press, Cambridge, 1987)

\bibitem{duval82}
P.D. Duval, B.R. Barrett, Nucl. Phys. A \textbf{376}, 213  (1982)

\bibitem{Maya2022}
E.~Maya-Barbecho, J.E. Garc\'{\i}a-Ramos, Phys. Rev. C \textbf{105}, 034341 (2022)

\bibitem{Maya2023}
E.~Maya-Barbecho, S.~Baid, J.M. Arias, J.E. Garc\'{\i}a-Ramos, Phys. Rev. C \textbf{108}, 034316 (2023)

\bibitem{Gavr19}
N.~Gavrielov, A.~Leviatan, F.~Iachello, Phys. Rev. C \textbf{99}, 064324 (2019)

\bibitem{Gavr22}
N.~Gavrielov, A.~Leviatan, F.~Iachello, Phys. Rev. C \textbf{105}, 014305 (2022)

\bibitem{Garc09}
J.~Garc\'{\i}a-Ramos, K.~Heyde, Nucl. Phys. A \textbf{825}, 39  (2009)

\bibitem{Bohr75}
A.~Bohr, B.R. Mottelson, \emph{{Nuclear Structure, Vol. II}} (W. A. Benjamin, Reading MA, 1975)

\bibitem{Garc20}
J.E. Garc\'{\i}a-Ramos, K.~Heyde, Phys. Rev. C \textbf{102}, 054333 (2020)

\end{thebibliography}
\end{document}